\documentclass[10pt, final, journal, letterpaper, twocolumn]{IEEEtran}

\usepackage{amsfonts}
\usepackage[dvips]{graphicx}
\usepackage{times}
\usepackage{cite}
\usepackage{amsmath}
\usepackage{cases}
\usepackage{array}
\usepackage{dsfont}
\usepackage{amssymb}

\usepackage{stfloats}
\usepackage{slashbox}
\usepackage{graphicx}
\usepackage{footnote}
\usepackage{booktabs}
\usepackage{array}
\usepackage{bm}
\usepackage{color}

\begin{document}

\title{Massive MIMO Multicasting in Noncooperative Cellular Networks}

\author{\IEEEauthorblockN{Zhengzheng Xiang, Meixia Tao, \IEEEmembership{Senior Member, IEEE}, and Xiaodong Wang, \IEEEmembership{Fellow, IEEE}}
\thanks{The material in this paper was presented in part at the IEEE International Conference on Communications (ICC), Sydney, Australia, June 10-14, 2014.}
\thanks{Z. Xiang and M. Tao are with the Department of
Electronic Engineering at Shanghai Jiao Tong University, Shanghai,
200240, P. R. China. Email: \{7222838, mxtao\}@sjtu.edu.cn.}\thanks{X. Wang is with the Department of Electrical Engineering at Columbia University, New York, USA. Email: wangx@ee.columbia.edu.}
\thanks{This work is supported by the National Natural Science Foundation of China under grants 61329101, 61322102, and 61221001.}
}
\maketitle

\begin{abstract}
We study physical layer multicasting in cellular networks where each base station (BS) is equipped with a very large number of antennas and transmits a common message using a single beamformer to multiple mobile users. The messages sent by different BSs are independent and the BSs do not cooperate. We first show that when each BS knows the perfect channel state information (CSI) of its own served users, the asymptotically optimal beamformer at each BS is a linear combination of the channel vectors of its multicast users. Moreover, the optimal and explicit combining coefficients are obtained. Then we consider the imperfect CSI scenario where the CSI is obtained through uplink channel estimation in time-division duplex systems. We propose a new pilot scheme that estimates the composite channel which is a linear combination of the individual channels of multicast users in each cell. This scheme is able to completely eliminate pilot contamination. The pilot power control for optimizing the multicast beamformer at each BS is also derived. Numerical results show that the asymptotic performance of the proposed scheme is close to the ideal case with perfect CSI. Simulation also verifies the effectiveness of the proposed scheme with finite number of antennas at each BS.

\end{abstract}

\begin{IEEEkeywords}
Physical layer multicasting, massive MIMO, asymptotic orthogonality, pilot contamination.
\end{IEEEkeywords}

\section{Introduction}
The array of emerging wireless data services, such as media streaming, cell broadcasting, and mobile TV have been propelling the development of new wireless communication technologies to make efficient use of the precious spectrum resources. One typical scenario is the delivery of common information, such as headline news, financial data, or  hot video, to multiple mobile users. Enabling such service is the wireless multicasting technique that transmits the same information to a group of users simultaneously. Wireless multicasting has been included in  the Third Generation Partnership Project (3GPP) LTE standards known as evolved Multimedia Broadcast Multicast Service (eMBMS) \cite{Lecompte}. Recently, mobile operators Verizon and AT\&T announced their plans to build LTE-multicast/broadcast networks for more efficient content delivery to  a large number of users \cite{Verzion,ATT}.

By exploiting the channel state information (CSI) at the transmitter equipped with multiple antennas, physical layer multicasting in the form of beamforming is a promising way to improve the wireless multicast throughput. The multicast beamforming was first considered in \cite{Sidiropoulos,Sunyan,Chenbiao} for a single group of users. The similar problem of multicasting to multiple cochannel groups was then studied in \cite{Karipidis,Gaoyan}. In \cite{Xiang}, the coordinated multicast beamforming in multicell networks was investigated. The authors in \cite{Ozbek} studied the multicast beamforming and resource allocation for orthogonal frequency division multiplexing (OFDM) systems. In \cite{Zhuhao}, the multicast precoding was considered for a single-group network.  Different from the traditional unicast beamforming problem in multi-user or multi-cell systems, the multicast beamforming problem is generally NP-hard \cite{Sidiropoulos} and hence more challenging. A key tool for finding near-optimal solutions in \cite{Sidiropoulos,Karipidis,Gaoyan,Xiang,Ozbek} is semidefinite relaxation (SDR). Other approaches such as channel orthogonalization-based beamforming, stochastic beamforming and space-time coding-based beamforming were considered in \cite{Hunger,Abdelkader,Wusisi,Wangjianqi,Wuxiaoxiao,WenXin}.

Recently, the massive, or large-scale multiple-input multiple output (MIMO) technology has emerged as a promising and efficient approach to mitigating the severe cochannel interference caused by the ever-increasing mobile data traffic \cite{Rusek}. It was first shown in \cite{Marzetta} that by applying a large number of antennas at each base station (BS) in a noncooperative cellular network, the intracell interference can be substantially reduced without sophisticated beamformer design. Moreover the transmit power at the BS can be made arbitrarily small. Other related works on the massive MIMO technology can be found, for example, in \cite{Larsson1,Marzetta1,Debbah,Alrabadi}. Massive MIMO provides a new dimension that can be exploited to further increase the throughput of wireless networks and has become an important candidate of key technologies for the future 5G wireless communication systems \cite{JianzhongZhang1,JianzhongZhang3,JianzhongZhang2}.

The goal of this paper is to investigate massive MIMO in the context of multicast transmission in cellular networks. In urban areas, where wireless multicasting finds applications, there are typically many high buildings with mobile users located at different floors. For such scenario, BSs with massive MIMO can efficiently shape the transmitted signals and significantly increase the quality of service compared with traditional BSs \cite{Larsson2}. On the other hand, coordinated transmission among multiple cells can generally provide better performance \cite{Xiang,Dahrouj,Zhangrui,Wangxiaodong}. However, for massive MIMO systems, due to the large amount of CSI information associated with the large number of BS antennas, coordination among different cells will induce a huge amount of information exchange overhead, which is impractical for systems with limited-capacity backhaul links. Moreover, as shown in \cite{Marzetta,Marzetta2}, for massive MIMO networks, coordinated transmission becomes less necessary compared with conventional systems since simple signal processing techniques can efficiently mitigate interference. Thus, in this paper, we will focus on noncooperative transmission in multicell multicast networks with massive MIMO. Note that the aforementioned works on MIMO multicast are all for traditional finite-size MIMO and they mainly used numerical optimization tools to obtain the beamformers except \cite{Hunger,Abdelkader}.

One challenge associated with massive MIMO is the acquisition of CSI at the transmitter, especially for the downlink transmission. Currently, the common assumption is the TDD protocol where the BS estimates the uplink channels and obtains the downlink CSI by exploiting the channel reciprocity. However, this approach suffers from the so-called pilot contamination problem in the multicell scenario since the users in other cells may use the same (or nonorthogonal) pilot sequences causing non-vanishing interference to the channel estimate at the BS, and the performance of the network will be severely degraded. A few initial attempts have been made very recently to tackle this problem. For example, in \cite{Marzetta2}, a time-shifted scheme was proposed to suppress the effects of pilot contamination. In \cite{Gesbert}, the coordinated channel estimation for massive MIMO networks was proposed by exploiting the second-order statistical information about the user channels. These works, however, are not specifically designed for multicast transmission.

The main contributions of this paper are as follows. First, we consider the multicast beamformer design with perfect CSI. We show that the asymptotically optimal beamforming vector for each BS is a linear combination of the channel vectors of the multicast users served in its cell. The optimal combining coefficients are obtained and they explicitly depend on the large-scale attenuation of the channels.

Second, we consider the imperfect CSI scenario using the conventional channel estimation method, and characterize the impact of pilot contamination on the signal-to-interference-plus-noise ratio (SINR) of each user.

Third and most importantly, we propose a novel pilot scheme for multicell multicasting to tackle the pilot contamination problem. The key idea is to assign one pilot sequence to one cell which is shared by all the multicast users served in the cell, and orthogonal pilot sequences are assigned to different cells. In this way, each BS estimates a composite channel which is a linear combination of the individual channels of its served multicast users. Such a composite channel has the same structure as the asymptotically optimal multicast beamformer and hence can serve as the beamfomer. By applying pilot power control among the users in each cell, we can optimize the composite channel and hence the multicast beamformer at each BS. The closed-form pilot power control rule is obtained and the SINR gaps between the perfect CSI case and the proposed pilot scheme with power control are analytically characterized. Compared with the traditional pilot scheme which has an SINR ceilling due to pilot contamination, our proposed pilot scheme completely eliminates pilot contamination and has a constant gap to the perfect CSI case. This gap is quite small and is independent of the BS transmitting power. Besides, we consider the synchronization issue for the pilot sequence transmission and analyze the effect of asynchrony on the performance of our proposed scheme.

The remainder of the paper is organized as follows. In Section II, the system model for multicell multicasting with massive MIMO is presented. Section III considers the beamformer design with perfect CSI. In Section IV, we consider the imperfect CSI scenario and  analyze the negative effects of the pilot contamination. In Section V, a novel pilot scheme for multicast in multicell is proposed which completely eliminates pilot contamination and the optimal pilot power control is derived. Section VI provides numerical and simulation results. Concluding remarks are made in Section VII.

{\sl Notations}: Boldface uppercase letters denote matrices and boldface lowercase letters denote vectors. $\mathds{R}$, $\mathds{C}$ and $\mathds{Z}^{+}$ denote the sets of real numbers, complex numbers, and positive integers, respectively. $(\cdot)^\textsl{T}$, $(\cdot)^\textsl{*}$, $(\cdot)^\textsl{H}$ and $\mbox{Tr}\{\cdot\}$ are the transpose, conjugate, Hermitian transpose and trace operators, respectively. $\mathds{E}(\cdot)$ is the expectation operator. ${\bf I}_N$ denotes the $N\times N$ identity matrix. ${\bf x}\sim \mathcal{CN}({\bf 0},{\bf I})$ means ${\bf x}$ is a circular symmetric complex Gaussian random vectors whose mean vector is ${\bf 0}$ and covariance matrix is ${\bf I}$. $f(y)\sim o(y)$ means that $\lim\limits_{y\rightarrow \infty}\frac{f(y)}{y}=0$.

\section{System Model}
We consider the downlink transmission of a multicast cellular network comprising $N$ cells and $K$ mobile users per cell as shown in Fig. $1$. The BS in each cell is equipped with $M$ antennas where $M$ is a very large number under massive MIMO, and every mobile user has a single antenna. The channel between each BS and each user is assumed to be flat and quasi-static fading. Let ${\bf g}_{i,j,k}^H$ denote the $1\times M$ channel vector from the BS in the $i$th cell to the $k$th user in the $j$th cell. It can be written as
\begin{eqnarray}
{\bf g}_{i,j,k}^H=\sqrt{\beta_{i,j,k}}{\bf h}_{i,j,k}^H,
\end{eqnarray}
where $\{\beta_{i,j,k}\in \mathds{R}\}$ are the large-scale channel attenuations which change slowly and can be tracked easily, and $\left\{{\bf h}_{i,j,k}\sim \mathcal{CN}({\bf 0},{\bf I}_M)\right\}$ are the small-scale fading coefficients and modeled as independent and identically distributed (i.i.d.) random vectors. Let $s_i$ denote the multicast information symbol for all the $K$ users in the $i$th cell with $\mathds{E}\left[|s_i|^2\right] = 1$ and ${\bf w}_i$ be the corresponding $M\times 1$ unit-norm multicast beamforming vector. The transmitted signal from BS $i$ is then given by
\begin{eqnarray}
{\bf x}_i=\sqrt{p_i}{\bf w}_is_i,
\end{eqnarray}
where $p_i$ is the transmit power by BS $i$. We assume that all the BSs and users are perfectly synchronized in time and use the same spectrum band. The discrete-time baseband signal received by the $k$th user in the $i$th cell is given by, for $k=1,..., K$ and $i=1,...,N$,
\begin{eqnarray}\nonumber
y_{i,k}=\sum\limits_{j=1}^N{\bf g}_{j,i,k}^H{\bf x}_j+z_{i,k}~~~~~~~~~~~~~~~~~~~~~~~~~~~~\\\label{y_i_k}
=\sqrt{p_i}{\bf g}_{i,i,k}^H{\bf w}_is_i+\sum\limits_{j\neq i}^{N}\sqrt{p_j}{\bf g}_{j,i,k}^H{\bf w}_js_j+z_{i,k}
\end{eqnarray}
where $z_{i,k}\sim \mathcal{CN}(0,\sigma^2)$ is the additive noise. In \eqref{y_i_k}, the second term on the right-hand side (RHS) is the interference from the BSs in other cells. Based on the received signal model in \eqref{y_i_k}, the performance of each user can be characterized by the output SINR, defined as
\begin{eqnarray}\label{sinr}
{\mbox{SINR}_{i,k}}=\frac{p_i|{\bf g}_{i,i,k}^H{\bf w}_i|^2}{\sum\limits_{j\neq i}^Np_j|{\bf g}_{j,i,k}^H{\bf w}_j|^2+\sigma^2}.
\end{eqnarray}
Since a common message is transmitted to the $K$ users in each cell, we characterize the performance of each cell by the user with the minimum SINR.

\begin{figure}
\begin{centering}
\includegraphics[scale=0.85]{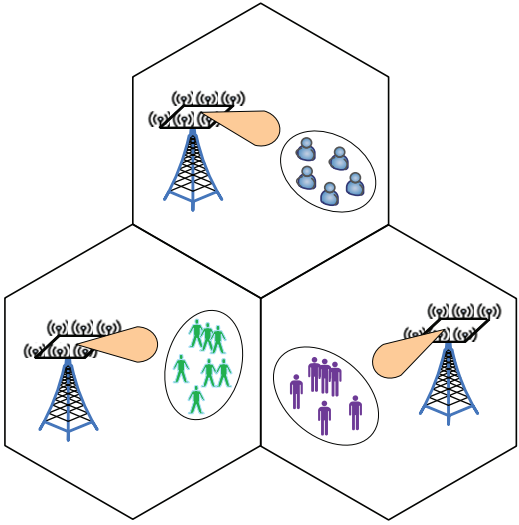}
\vspace{-0.1cm} \caption{Multicell multicast network with massive MIMO.}
\label{fig:multicell_multicast_system}
\end{centering}
\vspace{-0.3cm}
\end{figure}

Before moving on to the next section, we review two basic results on random vectors \cite{Cramer} that will be useful later on. Let ${\bf x}=\left[x_1,...,x_n\right]^T$ and ${\bf y}=\left[y_1,...,y_n\right]^T$ be two mutually independent $n\times 1$ vectors whose elements are i.i.d zero-mean random variables with variances $\sigma_x^2$ and $\sigma_y^2$, respectively. Then from the law of large numbers, we have
\begin{eqnarray}\label{asymp_opt}
\lim\limits_{n\rightarrow \infty}\frac{{\bf x}^H{\bf x}}{n}\overset{\text{a.s.}}{\rightarrow}\sigma_x^2,~\mbox{and}~\lim\limits_{n\rightarrow \infty}\frac{{\bf x}^H{\bf y}}{n}\overset{\text{a.s.}}{\rightarrow}0.
\end{eqnarray}
Here, $\overset{\text{a.s.}}{\rightarrow}$ denotes the almost sure convergence.

\section{Asymptotic Analysis with Perfect CSI }
In this section, we provide the asymptotic analysis for the multicell multicast network with massive MIMO when each BS knows the perfect CSI of its served users. Since for multicast transmission, the performance of the worst user in each cell is the bottleneck, our goal of the beamformer design for each BS is to maximize the minimum $\mbox{SINR}$ of the $K$ served users. This is formulated as the following optimization problem
\begin{eqnarray}\label{maxmin_opt}
{\cal P}: ~\max\limits_{\{{\bf w}_i\}}~\min\limits_{\forall k}~~\frac{p_i|{\bf g}_{i,i,k}^H{\bf w}_i|^2}{\sum\limits_{j\neq i}^Np_j|{\bf g}_{j,i,k}^H{\bf w}_j|^2+\sigma^2}\\\nonumber\mbox{s.t.}~~~~\|{\bf w}_i\|=1.~~~~~~~~~~~~~~~
\end{eqnarray}

\subsection{Asymptotically Optimal Beamformer Structure}
The optimization problem ${\cal P}$ for each BS is equivalent to the conventional single-cell multicast beamforming problem in \cite{Sidiropoulos}. Due to the NP-hardness, its optimal solution cannot be obtained efficiently in general. In our case, since each BS has a very large number of antennas, the different channels become asymptotically orthogonal. Therefore, we are able to obtain the asymptotically optimal structure of the beamforming vector in the following theorem.

$\mbox{{\bf Theorem 1}}$: The asymptotically optimal beamformer for each BS when $M\rightarrow\infty$ is a linear combination of the channels between the BS and its served users, i.e.,
\begin{eqnarray}\label{w_i2}
{\bf w}_i=\sum\limits_{k=1}^K\xi_{i,k}{\bf g}_{i,i,k},~\forall i\in \{1,...,N\},
\end{eqnarray}
where $\{\xi_{i,k}\}$ are the linear combination coefficients.
\begin{proof}
The proof is based on the asymptotic orthogonality of the channel vectors when $M$ is large. We first show that with the beamformers in \eqref{w_i2}, the intercell interference that each user suffers from diminishes as $M \to \infty$:
\begin{eqnarray}\nonumber
y_{i,k}=\sqrt{p_i}{\bf g}_{i,i,k}^H{\bf w}_is_i+\sum\limits_{j\neq i}^{N}\sqrt{p_j}{\bf g}_{j,i,k}^H{\bf w}_js_j+z_{i,k}~\\\nonumber=\sqrt{p_i}{\bf g}_{i,i,k}^H\left(\sum\limits_{k'=1}^K\xi_{i,k'}{\bf g}_{i,i,k'}\right)s_i~~~~~~~~~~~~~~~\\\nonumber+\sum\limits_{j\neq i}^{N}\sqrt{p_j}{\bf g}_{j,i,k}^H\left(\sum\limits_{k'=1}^K\xi_{j,k'}{\bf g}_{j,j,k'}\right)s_j+z_{i,k}\\=\underbrace{\sqrt{p_i}\xi_{i,k}{\bf g}_{i,i,k}^H{\bf g}_{i,i,k}s_i}_{\Gamma}+o(\Gamma)+z_{i,k}.~~~~~~~~~~~
\end{eqnarray}

Next if it does not have this structure, the beamformer of BS $i$ can be written as
\begin{eqnarray}\label{th1}
{\bf w}'_i=\sum\limits_{k=1}^K\xi'_{i,k}{\bf g}_{i,i,k}+\sum\limits_{t=1}^{M-K}\zeta_{i,t}{\bf u}_{i,t},
\end{eqnarray}
where $\{{\bf u}_{i,t}\}_{t=1}^{M-K}$ is an orthonormal basis for the orthogonal complement of the space spanned by $\{{\bf g}_{i,i,k}\}_{k=1}^{K}$.

From \eqref{th1} it is seen that the component $\sum\limits_{t=1}^{M-K}\zeta_{i,t}{\bf u}_{i,t}$ makes no contribution to the useful signals of the users in the $i$th cell but will cause extra interference to the users in the other cells. Hence, we can always reconstruct another beamformer for this BS without changing the transmit power as
\begin{eqnarray}
{\bf w}''_i=\sum\limits_{k=1}^K\eta\xi'_{i,k}{\bf g}_{i,i,k}.
\end{eqnarray}
Here,
\begin{eqnarray}\nonumber
\eta=\sqrt{\left(\sum\limits_{k=1}^K|\xi'_{i,k}|^2+\sum\limits_{t=1}^{M-K}|\zeta_{i,t}|^2\right){\big/\left(\sum\limits_{k'=1}^K|\xi'_{i,k}|^2\right)}}>1.
\end{eqnarray}
Given that the beamformers of other BSs keep unchanged, the received power of the users in the $i$th cell under ${\bf w}''_i$ is larger than that under ${\bf w}'_i$, and the interference caused by ${\bf w}''_i$ to the users in other cells still vanishes. This completes the proof.
\end{proof}

From Theorem $1$, we can see that the massive MIMO beamformer for the multicell multicast network has a rather simple structure. In this paper, we will exploit this structure to obtain the optimal multicast beamformers under  both perfect and estimated CSI.

Next, we will derive the asymptotic SINR achieved at the user receiver for the considered network. We first normalize the beamforming vector for each BS $i$ as
\begin{eqnarray}
{\bf w}_i=\frac{\sum\limits_{k=1}^K\xi_{i,k}{\bf g}_{i,i,k}}{\alpha_i\sqrt{M}}
\end{eqnarray}
where the normalization factor is
\begin{eqnarray}
\alpha_i=\frac{\left\|\sum\limits_{k=1}^K\xi_{i,k}{\bf g}_{i,i,k}\right\|}{\sqrt{M}}.
\end{eqnarray}
According to \eqref{asymp_opt}, we derive the asymptotic behavior of ${\alpha_i^2}$ as below
\begin{eqnarray}\nonumber
\lim\limits_{M\rightarrow \infty}{\alpha_i^2}=\lim\limits_{M\rightarrow \infty}\frac{\left(\sum\limits_{k=1}^K\xi_{i,k}{\bf g}_{i,i,k}\right)^H\left(\sum\limits_{k=1}^K\xi_{i,k}{\bf g}_{i,i,k}\right)}{M}\\\nonumber=\lim\limits_{M\rightarrow \infty}\frac{\sum\limits_{k=1}^K\sum\limits_{k'=1}^K\left|\xi_{i,k}^*\xi_{i,k'}\right|\sqrt{\beta_{i,i,k}\beta_{i,i,k'}}{\bf h}_{i,i,k}^H{\bf h}_{i,i,k'}}{M}\\\label{alpha_i}=\sum\limits_{k=1}^K\left|\xi_{i,k}\right|^2\beta_{i,i,k}.~~~~~~~~~~~~~~~~~~~~~~~~~~~~~~~~~~~~~~~~
\end{eqnarray}

Assume that the transmit power of each BS $p_i$ is scaled with $M$ according to $p_i=\frac{E_i}{M}$ where $E_i$ is fixed. Then, the asymptotic SINR of user $k$, for $k=1,..,K$ at cell $i$ is derived as
\begin{eqnarray}\nonumber
\lim\limits_{M\rightarrow\infty}\mbox{SINR}_{i,k}=\lim\limits_{M\rightarrow\infty}\frac{p_i|{\bf g}_{i,i,k}^H{\bf w}_i|^2}{\sum\limits_{j\neq i}^Np_j|{\bf g}_{j,i,k}^H{\bf w}_j|^2+\sigma^2}~~~~~~~~~\\\label{th2_SINR}=\lim\limits_{M\rightarrow\infty}\frac{\frac{E_i\beta_{i,i,k}}{\alpha_i^2}\left|\frac{\sum\limits_{k'=1}^K\xi_{i,k'}\sqrt{\beta_{i,i,k'}}{\bf h}_{i,i,k}^H{\bf h}_{i,i,k'}}{M}\right|^2}{\sum\limits_{j\neq i}^N\frac{E_j\beta_{j,i,k}}{\alpha_j^2}\left|\frac{\sum\limits_{k'=1}^K\xi_{j,k'}\sqrt{\beta_{j,j,k'}}{\bf h}_{j,i,k}^H{\bf h}_{j,j,k'}}{M}\right|^2+\sigma^2}.
\end{eqnarray}
Applying \eqref{asymp_opt} and plugging \eqref{alpha_i} into \eqref{th2_SINR}, we obtain
\begin{eqnarray}\label{asym_SINR1}
\lim\limits_{M\rightarrow\infty}\mbox{SINR}_{i,k}=\frac{\lambda_{i,k}E_i\beta_{i,i,k}}{\sigma^2},
\end{eqnarray}
where
\begin{eqnarray}\label{th2_1}
\lambda_{i,k}=\frac{\left|\xi_{i,k}\right|^2\beta_{i,i,k}}{\sum\limits_{k'=1}^K\left|\xi_{i,k'}\right|^2\beta_{i,i,k'}}.
\end{eqnarray}

From \eqref{asym_SINR1} it is seen that with a large $M$, each BS just needs the power $\frac{E_i}{M}$ to achieve the same performance as that of a single-input single-output (SISO) system with transmit power $\lambda_{i,k}E_i$, without any interference and the fast-fading effect, which coincides with \cite{Larsson1}. For each antenna, its transmit power will be $\frac{E_i}{M^2}$. Thus, given the SINR target, we can reduce the power consumption at the BS by increasing the BS antennas. In the rest of this paper, we will assume that $p_i=\frac{E_i}{M},\forall i$.

\subsection{Optimal Beamforming Parameters}
In this subsection, we will derive the optimal linear combination coefficients for the asymptotically optimal beamformer in Theorem $1$ based on the max-min criterion in \eqref{maxmin_opt}. Recall that the asymptotic SINR is directly related to $\{\lambda_{i,k}\}$ instead of $\{\xi_{i,k}\}$ as defined in \eqref{th2_1}. Moreover, an important observation is that
\begin{eqnarray}\label{ob_lamda}
\sum\limits_{k=1}^K\lambda_{i,k}=1.
\end{eqnarray}
Therefore, optimizing the coefficients $\{\xi_{i,k}\}$ is equivalent to finding the optimal $\{\lambda_{i,k}\}$. This problem can be formulated as
\begin{eqnarray}\label{prob_perfect_CSI}
{\cal Q}: ~\max\limits_{\{\lambda_{i,k}\}}~\min\limits_{\forall k}~~\frac{\lambda_{i,k}E_i\beta_{i,i,k}}
{\sigma^2}\\\nonumber\mbox{s.t.}~~~~~~~~\eqref{ob_lamda}.~~~~~~~~
\end{eqnarray}

$\mbox{{\bf Lemma 2}}$: The optimal solution of problem ${\cal Q}$ is obtained when all the $K$ users have the same SINR, i.e.,
\begin{eqnarray}\label{th3_1}
\frac{\lambda_{i,1}E_i\beta_{i,i,1}}{\sigma^2}=...=\frac{\lambda_{i,K}E_i\beta_{i,i,K}}{\sigma^2}
\end{eqnarray}
and the optimal $\{\lambda_{i,k}^*\}$ are
\begin{eqnarray}\label{th3_2}
\lambda_{i,k}^*=\frac{1}{\sum\limits_{k'=1}^K\frac{\beta_{i,i,k}}{{\beta_{i,i,k'}}}},~\forall i,~\forall k.
\end{eqnarray}
\begin{proof}
Please refer to Appendix A.
\end{proof}

According to Lemma $2$ and (\ref{th2_1}) and after some manipulations, we can obtain the following theorem:

$\mbox{{\bf Theorem 3}}$: The asymptotically optimal multicast beamforming vector is given by
\begin{eqnarray}\label{w_i1}
{\bf w}_i^*=\mu_i\sum\limits_{k=1}^K\frac{{\bf g}_{i,i,k}}{\beta_{i,i,k}}, ~\forall i
\end{eqnarray}
where the common factor $\mu_i$ is
\begin{eqnarray}
\mu_i=\frac{1}{\sqrt{M\sum\limits_{k=1}^K\frac{1}{\beta_{i,i,k}}}},
\end{eqnarray}
and the resulting asymptotic SINR of each user is given by
\begin{eqnarray}\label{asym_SINR3}
\lim\limits_{M\rightarrow\infty}\mbox{SINR}_{i,k}=\frac{E_i}{\sigma^2\sum\limits_{k'=1}^K\frac{1}{\beta_{i,i,k'}}}, ~\forall i,k.
\end{eqnarray}

From Theorem $3$ it is seen that the optimal combination coefficients in the beamformer depend only on the large-scale channel parameters $\{\beta_{i,i,k} \}$, and the user with the larger large $\beta_{i,i,k}$ will have lower weight in the beamformer. Also, when the user number $K$ increases, the performance will decrease due to the multicast feature of the network.

Substituting the asymptotically optimal ${\bf w}_i^*$ \eqref{w_i1} into the received signal model in \eqref{y_i_k}, we obtain that
\begin{eqnarray}\nonumber
y_{i,k}=\sqrt{\frac{E_i}{M}}\sum\limits_{k'=1}^K\frac{\mu_i{\bf g}_{i,i,k}^H{\bf g}_{i,i,k'}}{\beta_{i,i,k'}}s_i~~~~~~~~~~~~~~~~\\\nonumber+\sum\limits_{j\neq i}^{N}\sqrt{\frac{E_j}{M}}\sum\limits_{k'=1}^K\frac{\mu_j{\bf g}_{j,i,k}^H{\bf g}_{j,j,k'}}{\beta_{j,j,k'}}s_j+z_{i,k}\\=\underbrace{\sqrt{\frac{E_i\mu_i^2}{M\beta_{i,i,k}}}{\bf g}_{i,i,k}^H{\bf g}_{i,i,k}s_i}_{\Gamma}+o(\Gamma)+z_{i,k}.~~~
\end{eqnarray}
For large $M$, the effective channel coefficient for the $k$th user in the $i$th cell converges to:
\begin{eqnarray}\nonumber
\sqrt{\frac{E_i\mu_i^2}{M\beta_{i,i,k}}}{\bf g}_{i,i,k}^H{\bf g}_{i,i,k}\rightarrow\sqrt{\frac{E_i}{\sum\limits_{k'=1}^K\frac{\beta_{i,i,k}^2}{\beta_{i,i,k'}}}}.
\end{eqnarray}
Thus, to decode its desired message, each mobile user does not need to know the small-scale fading channel vector which has a big size and changes rather quickly, but it only needs to learn the effective channel which is a scalar and only contains the slowly-changing large-scale fading parameters. Such feature is preferable for massive MIMO networks since the acquisition of CSI at the user end is highly challenging.

The asymptotic behavior of multicasting was considered in \cite{Jindal,Seung,Kim} when the number of transmit antennas goes to infinity. However, the scopes of those works are different from ours. In particular, in \cite{Jindal,Seung}, the capacity scaling of the multicast network is analyzed when the number of transmit antennas grows large, without considering the asymptotic beamformer structure. In \cite{Kim}, the optimal beamformer is given for the two-user case, which cannot be generalized to the case of an arbitrary number of users. In our work, we focus on multicell massive MIMO and derive the asymptotically optimal beamformer structure for multiuser multicast. In the special case with two users, our beamformer structure coincides with the solution of \cite{Kim} with the same form of linear combination.

Thus far, we have obtained the asymptotically optimal beamformer for the massive MIMO multicasting when having perfect CSI. In the next section, we will consider the imperfect CSI scenario and investigate the pilot contamination issue for this network.

\section{Analysis with Imperfect CSI and Pilot Contamination}
For the downlink transmission of cellular networks, there are two main approaches for the acquisition of CSI at the BS. In frequency-division duplex (FDD) systems, each user first estimates the downlink channel and then feeds back some quantized version to the BS. In TDD systems, the user transmits the pilot signal to the BS. The BS estimates the uplink channel and treats it as the downlink CSI according to channel reciprocity, and no CSI feedback is needed.

In the massive MIMO network, since the size of the channel vector is very large, both the channel estimation and CSI feedback at the user end become very difficult and expensive. However, for channel estimation at the BS end, the time required for pilot transmission is independent of the number of BS antennas and it is much easier for the BS to do channel estimation. Thus, we assume the TDD mode and use uplink pilots in this paper. We also assume that the channel varies slowly enough such that the channel reciprocity holds.

We  consider the estimation of individual channel vectors. Specifically, the same set of $K$ orthogonal pilot sequences is used in all the $N$ cells. Every $k$th user in all the cells uses the same $1\times \tau$ pilot sequence
\begin{eqnarray}\nonumber
\sqrt{\tau}{\bm \phi}_k=\sqrt{\tau}\left[\phi_{k1},...,\phi_{k\tau}\right],
\end{eqnarray}
where $\tau\geq K$ and ${\bm \phi}_k{\bm \phi}_{k'}^H=1$ for $k=k'$ and $0$ otherwise.

At the pilot transmission phase, all the users simultaneously transmit the pilot signals and the $i$th BS receives the pilot signal as
\begin{eqnarray}
{\bf Y}_{B_i}=\sum\limits_{l=1}^N\sum\limits_{k=1}^K\sqrt{p_u\tau}{\bf g}_{i,l,k}{\bm \phi}_{k}+{\bf Z}_{B_i},
\end{eqnarray}
where $p_u$ is the average transmit power of the users and ${\bf Z}_{B_i}\in \mathds{C}^{M\times\tau}$ is the additive noise matrix with each element being a symmetric complex Gaussian random variable with mean zero and variance $\sigma_p^2$. The BS then estimates the channel vector for the $k$th user as
\begin{eqnarray}\nonumber
\hat{{\bf g}}_{i,i,k}&=&{\bf Y}_{B_i}{\bm \phi}_k^H \\\nonumber
&=&\sum\limits_{l=1}^N\sum\limits_{k'=1}^K\sqrt{p_u\tau}{\bf g}_{i,l,k'}{\bm \phi}_{k'}{\bm \phi}_k^H+{\bf Z}_{B_i}{\bm \phi}_k^H\\
\label{CSI_estimate}
&=&\sum\limits_{l=1}^N\sqrt{p_u\tau}{\bf g}_{i,l,k}+{\bf z}_{B_{ik}},
\end{eqnarray}
where ${\bf z}_{B_{ik}}={\bf Z}_{B_i}{\bm \phi}_k^H\sim\mathcal{CN}({\bf 0},\sigma_p^2{\bf I}_M)$. It is seen that the channel estimate at each BS not only contains the desired CSI $\sqrt{p_u\tau}{\bf g}_{i,i,k}$, but also the undesired CSI $\sum_{l\neq i}^N\sqrt{p_u\tau}{\bf g}_{i,l,k}$. That is, the channel estimation is contaminated by the interference from other cells.

Based on the estimated CSI in \eqref{CSI_estimate} and the beamformer structure in Theorem $1$, the normalized beamforming vector for BS $i$ can be written as
\begin{eqnarray}\label{w_i}
{\bf w}_i=\frac{\sum\limits_{k=1}^K\xi_{i,k}\hat{{\bf g}}_{i,i,k}}{\gamma_i\sqrt{M}},
\end{eqnarray}
where $\gamma_i$ is a normalization factor such as that $\|{\bf w}_i\|=1$.

As in Section III, we first derive the asymptotic behavior of $\gamma_i^2$ which will be used in the derivation of the asymptotic SINR:

\begin{eqnarray}\nonumber
\lim\limits_{M\rightarrow \infty}\gamma_i^2~~~~~~~~~~~~~~~~~~~~~~~~~~~~~~~~~~~~~~~~~~~~~~~~\\\nonumber=\lim\limits_{M\rightarrow \infty}\frac{\left\|\sum\limits_{k=1}^K\xi_{i,k}\left(\sum\limits_{l=1}^N\sqrt{p_u\beta_{i,l,k}\tau}{\bf h}_{i,l,k}+{\bf z}_{B_{ik}}\right)\right\|^2}{M}\\=\sum\limits_{k=1}^K\sum\limits_{l=1}^N\xi_{i,k}^2p_u\tau\beta_{i,l,k}+\sigma_p^2\sum\limits_{k=1}^K\xi_{i,k}^2\triangleq \gamma_{i,\infty}^2.~~~~~~~~~
\end{eqnarray}
Then, the asymptotic SINR of each user for the imperfect CSI scenario is given by:
\begin{eqnarray}\nonumber
\lim\limits_{M\rightarrow \infty}\mbox{SINR}_{i,k}=\lim\limits_{M\rightarrow\infty}\frac{p_i|{\bf g}_{i,i,k}^H{\bf w}_i|^2}{\sum\limits_{j\neq i}^Np_j|{\bf g}_{j,i,k}^H{\bf w}_j|^2+\sigma^2}\\\nonumber=\lim\limits_{M\rightarrow \infty}\frac{p_i\left|{\bf g}_{i,i,k}^H\left(\frac{\sum\limits_{k'=1}^K\xi_{i,k'}\hat{{\bf g}}_{i,i,k'}}{\gamma_i\sqrt{M}}\right)\right|^2}{\sum\limits_{j\neq i}^Np_j\left|{\bf g}_{j,i,k}^H\left(\frac{\sum\limits_{k'=1}^K\xi_{j,k'}\hat{{\bf g}}_{j,j,k'}}{\gamma_j\sqrt{M}}\right)\right|^2+\sigma^2}~\\\label{SINR_contamination}=\frac{\frac{E_i}{\gamma_{i,\infty}^2}\beta_{i,i,k}^2\xi_{i,k}^2p_u\tau}{\sum\limits_{j\neq i}^N\frac{E_j}{\gamma_{j,\infty}^2}\beta_{j,i,k}^2\xi_{j,k}^2p_u\tau+\sigma^2}\triangleq \mathcal{A}.~~~~~~~~~~~~~~~
\end{eqnarray}
In the extreme case where $E_i = E \to \infty,\forall i$, we have
\begin{eqnarray}\nonumber
\lim\limits_{E\rightarrow \infty}\mathcal{A}=\frac{\beta_{i,i,k}^2\xi_{i,k}^2}{\sum\limits_{j\neq i}^N\frac{\gamma_{i,\infty}^2}{\gamma_{j,\infty}^2}\beta_{j,i,k}^2\xi_{j,k}^2}.
\end{eqnarray}
That is, the asymptotic SINR reaches a ceiling and will not increase with the transmit power of BS. This clearly indicates the effect of pilot contamination.

Note that, based on the SINR expression in \eqref{SINR_contamination}, it is difficult to obtain the asymptotically optimal parameters $\{\xi_{i,k}\}$ like the perfect CSI case for our considered noncooperative cellular network.

\section{A Contamination-free Pilot Scheme}
In order to avoid pilot contamination, a simple and straightforward approach is to assign a different and orthogonal pilot sequence to each user in the whole network. However, this will need a very long pilot sequence which becomes prohibitive in practice. In this section, we propose a new pilot scheme for the multicell multicast network with large scale antenna arrays to eliminate the pilot contamination.

\subsection{The New Pilot Scheme}
The key idea of our proposed pilot scheme is that we apply a set of $N$ orthogonal pilot sequences for the whole network and the $K$ users in each cell share the same pilot sequence. Thus, each BS will obtain an estimated composite channel of the served $K$ users.

For cell $i$, the $K$ users transmit the same $1\times \omega$ pilot sequence
\begin{eqnarray}\nonumber
\sqrt{\omega}{\bm \psi}_i=\sqrt{\omega}\left[\psi_{i1},...,\psi_{i\omega}\right]
\end{eqnarray}
where $\omega\geq N$ and ${\bm \psi}_i{\bm \psi}_j^H=1$ for $i=j$ and $0$ otherwise. The received signal at BS $i$ is then written as
\begin{eqnarray}
{\bf Y}_{B_i}=\sum\limits_{l=1}^N\sum\limits_{k=1}^K\sqrt{p_u\omega}{\bf g}_{i,l,k}{\bm \psi}_l+{\bf Z}_{B_i}.
\end{eqnarray}
BS $i$ then estimates the composite channel vector for the users in cell $i$ as
\begin{eqnarray}\nonumber
\hat{{\bf g}}_{i,c}={\bf Y}_{B_i}{\bm \psi}_i^H~~~~~~~~~~~~~~~~~~~~~~~~~~~~~~~~~\\\nonumber=\sum\limits_{l=1}^N\sum\limits_{k=1}^K\sqrt{p_u\omega}{\bf g}_{i,l,k}{\bm \psi}_l{\bm \psi}_i^H+{\bf Z}_{B_i}{\bm \psi}_i^H\\\label{composite_ch}=\sum\limits_{k=1}^K\sqrt{p_u\omega}{\bf g}_{i,i,k}+{\bf z}_{B_{i}}.~~~~~~~~~~~~~~~~
\end{eqnarray}

Clearly, the estimate of the composite channel vector is a linear combination of the channel vectors of the $K$ users from the desired cell (plus the additive noise), and does not contain the channels of the users from other cells. According to Theorem $1$, the asymptotically optimal beamfomer for each BS is also a linear combination of the channels between the BS and its served users. Therefore, the BS can design the beamforming vector as
\begin{eqnarray}\nonumber
{\bf w}_i=\frac{\hat{{\bf g}}_{i,c}}{\left\|\hat{{\bf g}}_{i,c}\right\|}~~~~~~~~~~~~~~~~~~~~~~~\\=\sum\limits_{k=1}^K\frac{\sqrt{p_u\omega}}{\left\|\hat{{\bf g}}_{i,c}\right\|}{\bf g}_{i,i,k}+\frac{{\bf z}_{B_{i}}}{\left\|\hat{{\bf g}}_{i,c}\right\|}.
\end{eqnarray}
As a result, there will be no intercell interference for the composite channel estimation scheme.

\subsection{Pilot Power Control}
From the previous subsection, it is seen that the proposed pilot scheme completely eliminates the pilot contamination. However, with the resulting composite channel estimation, the BS no longer has the flexibility to optimize the beamformer coefficients.

In this subsection, we propose power control among the $K$ users of each cell in the pilot transmission phase to optimize the coefficients in each BS's beamformer, and to improve the network performance.

Let $p_{i,k}$ denote the transmit power of user $k$ in BS $i$ in the pilot transmission phase and be subject to the peak power constraint $p_{i,k} < p_u$. By replacing $p_u$ with $p_{i,k}$, the composite channel vector in \eqref{composite_ch} can be rewritten as
\begin{eqnarray}
\hat{{\bf g}}_{i,c}=\sum\limits_{k=1}^K\sqrt{\omega\beta_{i,i,k}p_{i,k}}{\bf h}_{i,i,k}+{\bf z}_{B_{i}}.
\end{eqnarray}
Then the normalized beamforming vector is
\begin{eqnarray}\label{composite_w}
{\bf w}_i=\frac{\sum\limits_{k=1}^K\sqrt{\omega\beta_{i,i,k}p_{i,k}}{\bf h}_{i,i,k}+{\bf z}_{B_{i}}}{\mu_i\sqrt{M}},
\end{eqnarray}
where $\mu_i$ is a normalization factor and its limiting value is given by
\begin{eqnarray}\label{mu_perfect}
\lim\limits_{M\rightarrow \infty}\mu_i^2=\omega\sum\limits_{k=1}^K\beta_{i,i,k}p_{i,k}+\sigma_p^2\triangleq \mu_{i,\infty}^2.
\end{eqnarray}
With the above beamformer, we derive the asymptotic SINR of each user when $M\rightarrow \infty$ for the proposed composite channel estimation scheme as follows
\begin{eqnarray}\nonumber
\lim\limits_{M\rightarrow \infty}\mbox{SINR}_{i,k}~~~~~~~~~~~~~~~~~~~~~~~~~~~~~~~~~~~~~~~~~~~~~~~~~~~~~~~\\\nonumber=\lim\limits_{M\rightarrow \infty}\frac{p_i\beta_{i,i,k}\left|{\bf h}_{i,i,k}^H\frac{\left(\sum\limits_{k'=1}^K\sqrt{\omega\beta_{i,i,k'}p_{i,k'}}{\bf h}_{i,i,k'}+{\bf z}_{B_{i}}\right)}{\mu_i\sqrt{M}}\right|^2}{\sum\limits_{j\neq i}^Np_j\beta_{j,i,k}\left|{\bf h}_{j,i,k}^H\frac{\left(\sum\limits_{k'=1}^K\sqrt{\omega\beta_{j,j,k'}p_{j,k'}}{\bf h}_{j,j,k'}+{\bf z}_{B_{j}}\right)}{\mu_j\sqrt{M}}\right|^2+\sigma^2}\\\nonumber=\frac{E_i\beta_{i,i,k}^2\omega p_{i,k}}{\mu_{i,\infty}^2\sigma^2}~~~~~~~~~~~~~~~~~~~~~~~~~~~~~~~~~~~~~~~~~~~~~~~~~~~~~~~~~\\\nonumber=\frac{E_i}{\sigma^2}\cdot\frac{\beta_{i,i,k}^2p_{i,k}}
{\sum\limits_{k'=1}^K\beta_{i,i,k'}p_{i,k'}+\frac{\sigma_p^2}{\omega}}.~~~~~~~~~~~~~~~~~~~~~~~~~~~~~~~~~~\mbox{(36)}~~~
\end{eqnarray}
\addtocounter{equation}{1}
Based on the max-min criterion, we can formulate the pilot power control problem for each cell as follows:
\begin{eqnarray}\label{Ti_ob1}
{\cal T}: ~\max\limits_{\{p_{i,k}\}}~\min\limits_{\forall k}~~\frac{\beta_{i,i,k}^2p_{i,k}}{\sum\limits_{k'=1}^K\beta_{i,i,k'}p_{i,k'}+\frac{\sigma_p^2}{\omega}}~~~~~~~~~\\\nonumber\mbox{s.t.}~~~~p_{i,k}\leq p_u,~\forall k, ~~~~~~~~~~~~~~~~~~~~~
\end{eqnarray}
where $p_u$ denotes the peak transmit power of each user.

Changing the objective function of ${\cal T}$, we can rewrite problem ${\cal T}$ as
\begin{eqnarray}\label{Ti_ob2}
{\cal T}: ~\min\limits_{\{p_{i,k}\}}~\max\limits_{\forall k}~\frac{\sigma_p^2}{\omega}\beta_{i,i,k}^{-2}p_{i,k}^{-1}+\sum\limits_{k'=1}^K\beta_{i,i,k}^{-2}\beta_{i,i,k'}p_{i,k}^{-1}p_{i,k'}\\\nonumber\mbox{s.t.}~~~~p_{i,k}\leq p_u,~\forall k\in\{1,...,K\}.~~~~~~~~~~~
\end{eqnarray}

To further simplify the objective function of the above optimization problem, we introduce a slack variable $t$ and reformulate the problem as follows
\begin{eqnarray}\label{Ti_ob3}
{\cal O}: ~\min\limits_{\{p_{i,k},t\}}~~~t~~~~~~~~~~~~~~~~~~~~~~~~~~~~~~~~~~~~~~~~~~~~~~~~\\\nonumber\mbox{s.t.}~~\frac{\sigma_p^2}{\omega}\beta_{i,i,k}^{-2}p_{i,k}^{-1}t^{-1}+\sum\limits_{k'=1}^K\beta_{i,i,k}^{-2}\beta_{i,i,k'}p_{i,k}^{-1}p_{i,k'}t^{-1}\leq 1\\\nonumber p_{i,k}\leq p_u,~\forall k\in\{1,...,K\}.~~~~~~~~~~~~~~~~~~~~~~~~~
\end{eqnarray}
The above problem is a geometric programming problem. It can be transformed to a convex problem and solved optimally using the software package \cite{Boyd}. By exploiting the specific structure of the problem, we further show that a closed-form expression for the optimal solution can be obtained.

$\mbox{{\bf Lemma 4}}$: At the optimal solution of problem \eqref{Ti_ob1}, the pilot power of user $k^\star$ with the minimum channel gain $\beta_{i,i,k^\star}$ is $p_u$, i.e.,
\begin{eqnarray}\label{p_star}
p_{i,k^\star}=p_u.
\end{eqnarray}
\begin{proof}
Please refer to Appendix B.
\end{proof}

The above lemma tells us that the user with the minimum channel gain should transmit the pilot sequence with full power. Based on this result, we can derive the optimal closed-form solution of problem \eqref{Ti_ob1} in the following theorem

$\mbox{{\bf Theorem 5}}$: The optimal solution of problem \eqref{Ti_ob1} is obtained when the achieved SINRs of all the users are the same, and the optimal pilot power for each user is
\begin{eqnarray}\label{simplified}
p_{i,k}^*=\frac{\beta_{i,i,k^\star}^2p_u}{\beta_{i,i,k}^2}, ~\forall k,
\end{eqnarray}
where $\beta_{i,i,k^\star}$ is the minimum channel gain among the $K$ users.
%

\begin{proof}
We first prove that the pilot power of the user $j^\star (\neq k^\star)$ with the second minimum fading ratio $\beta_{i,i,j^\star}$ is
\begin{eqnarray}\label{th5-1}
p_{i,j^\star}=\frac{\beta_{i,i,k^\star}^2p_u}{\beta_{i,i,j^\star}^2}.
\end{eqnarray}
by contradiction.

If the optimal solution satisfies $p_{i,j^\star}>\frac{\beta_{i,i,k^\star}^2p_u}{\beta_{i,i,j^\star}^2}$, the SINR of user $j^\star$ will be larger than that of user $k^\star$ since the expression of each user's SINR has a common denominator $\sum\limits_{k'=1}^K\beta_{i,i,k'}p_{i,k'}+\frac{\sigma_p^2}{\omega}$. Then we can set the pilot power of user $j^\star$ as $p_{i,j^\star}=\frac{\beta_{i,i,k^\star}^2p_u}{\beta_{i,i,j^\star}^2}$. Thus, it can be easily seen that the objective value of problem \eqref{Ti_ob1} will increase since the common denominator $\sum\limits_{k'=1}^K\beta_{i,i,k'}p_{i,k'}+\frac{\sigma_p^2}{\omega}$ decreases, which contradicts with the optimality.

Else if $p_{i,j^\star}<\frac{\beta_{i,i,k^\star}^2p_u}{\beta_{i,i,j^\star}^2}$ holds in the optimal solution, we will have that
\begin{eqnarray}\label{th5-2}
p_{i,k}<\frac{\beta_{i,i,k^\star}^2p_u}{\beta_{i,i,j^\star}^2}, \forall k\in \mathcal{S}\backslash\{k^\star\}
\end{eqnarray}
where $\mathcal{S}=\{1,2,...,K\}$. The reason is similar to the proof of Lemma $4$ and we omit the details here. Under such condition, the user with the minimum SINR must lie in the set $\mathcal{S}\backslash\{k^\star\}$ since the SINR of user $j^\star$ is smaller than that of user $k^\star$.

Now, based on \eqref{th5-2}, we can always scale up the pilot power of all the $K-1$ users with $\eta>1$:
\begin{eqnarray}
p'_{i,k}=\eta p_{i,k}, \forall k\in \mathcal{S}\backslash\{k^\star\}
\end{eqnarray}
while still maintaining that $p'_{i,k}<\frac{\beta_{i,i,k^\star}^2p_u}{\beta_{i,i,j^\star}^2}, \forall k\in \mathcal{S}\backslash\{k^\star\}$. After this scaling, although the SINR of user $k^\star$ decreases, the SINR of user $j^\star$ is still smaller than that of user $k^\star$ since $p'_{i,j^\star}<\frac{\beta_{i,i,k^\star}^2p_u}{\beta_{i,i,j^\star}^2}$, which means that the minimum SINR still lies in the set $\mathcal{S}\backslash\{k^\star\}$. On the other hand, it can be seen that the SINR of each user in $\mathcal{S}\backslash\{k^\star\}$ increases. Thus, the minimum SINR increases which contradicts with the optimality. Finally, we must have \eqref{th5-1}.

Using similar argument, we can continue the proof for the remaining users and the details are omitted. Thus, we complete the proof of this theorem.
\end{proof}

From the above theorem, we can see that the proposed pilot transmit power control can be performed by each individual user in a distributed manner. The only information that needs to be exchanged among the users is the minimum channel gain $\beta_{i,i,k^\star}$.

Substituting the optimal pilot power control \eqref{simplified} into \eqref{composite_w}, we can obtain the beamforming vector for the BS as follows:
\begin{eqnarray}\nonumber
{\bf w}_i=\sum\limits_{k=1}^K\frac{{\bf g}_{i,i,k}+\frac{{\bf z}_{B_i}}{\sqrt{\omega p_{i,k}}}}{\sqrt{\sum\limits_{k'=1}^K\beta_{i,i,k'}\frac{p_{i,k'}}{p_{i,k}}+\frac{\sigma_p^2}{\omega p_{i,k}}}\sqrt{M}}\\\label{w_i3}=\sum\limits_{k=1}^K\frac{{\bf g}_{i,i,k}+\frac{{\bf z}_{B_i}}{\sqrt{\omega p_{i,k}}}}{\sqrt{\sum\limits_{k'=1}^K\frac{\beta_{i,i,k}^2}{\beta_{i,i,k'}}+\frac{\sigma_p^2}{\omega p_{i,k}}}\sqrt{M}}.~~~~~
\end{eqnarray}
By comparing \eqref{w_i3} with \eqref{w_i1}, one can see that the beamforming vector obtained by the optimal pilot power control is almost the same as the optimal beamforming vector of the perfect CSI case except for the noise terms $\frac{{\bf z}_{B_i}}{\sqrt{\omega p_{i,k}}}$ and $\frac{\sigma_p^2}{\omega p_{i,k}}$. Thus, effectively, the optimization of pilot power at each user side plays the similar role as the optimization of the linear combination coefficients at the BS side in the perfect CSI case.


In the following, we characterize the SINR gap between the proposed pilot scheme and the perfect CSI case. Plugging \eqref{simplified} into (36), we can obtain the asymptotic SINR for our proposed pilot scheme with optimal power control as
\begin{eqnarray}\label{asym_SINR2}
\lim\limits_{M\rightarrow \infty}\mbox{SINR}'_{i,k}=\frac{E_i}{\sigma^2}\cdot\frac{1}{\sum\limits_{k'=1}^K\frac{1}{\beta_{i,i,k'}}+\frac{\sigma_p^2}
{\omega\beta_{i,i,k^\star}^2p_u}}.
\end{eqnarray}
Comparing \eqref{asym_SINR3} and \eqref{asym_SINR2}, we then obtain the SINR gap
\begin{eqnarray}\nonumber
\triangle \mbox{SINR}=10\log_{10}\bigg(\frac{E_i}{\sigma^2\sum\limits_{k'=1}^K\frac{1}{\beta_{i,i,k'}}}\bigg)~~~~~~~~~~~~~~~~~~~\\\nonumber-10\log_{10}\bigg(\frac{E_i}{\sigma^2}\cdot\frac{1}{\sum\limits_{k'=1}^K\frac{1}{\beta_{i,i,k'}}+\frac{\sigma_p^2}
{\omega\beta_{i,i,k^\star}^2p_u}}\bigg)\\=10\log_{10}\bigg(1+\frac{\sigma_p^2}{\omega p_u\beta_{i,i,k^\star}^2\sum\limits_{k'=1}^K\frac{1}{\beta_{i,i,k'}}}\bigg).~~~~
\end{eqnarray}
This gap decreases as the peak pilot power $p_u$ increases, and is independent of the transmit power of the BS.

\emph{Remark 1}: In the proposed pilot scheme for composite channel estimation, the length of the pilot should not be less than $N$. Recall that if the individual channel estimation scheme is applied, the pilot length will be at least $K$. Similar to the user number $K$, the cell number $N$ (adjacent cells) will not be too large and is proper for practical implementations.

\emph{Remark 2}: In the optimal pilot power control, the knowledge of large-scale channel attenuation factors of all users in the same cell, i.e. $\{\beta_{i,i,k}\}$ should be known. In practice, the channel attenuation changes very slowly compared with the small-scale fading, and hence can be estimated in advance using off-line approaches.  On the other hand, if this knowledge is not available, we simply let each user transmit at its peak power, but the performance will decrease. However, from the simulation results, we will see that the proposed scheme without pilot power control still outperforms the individual channel estimation scheme.

\subsection{Discussion On Synchronization}

During the uplink pilot transmission phase, the users are not necessarily perfectly synchronized. In addition, the pilot signal coming from different users may arrive at the BS with different propagation delays. In this subsection, we discuss the synchronization and delay issues of the proposed composite channel estimation.

We first consider the case that the time difference between the pilot arrivals at the BS is so small and hence negligible compared with one symbol duration. The channel is then effectively phase shifted by the time delay which can be modeled as
\begin{eqnarray}
{\bf Y}_{B_i}=\sum\limits_{l=1}^N\sum\limits_{k=1}^K\sqrt{p_{l,k}\omega}{\bf g}_{i,l,k}e^{j\theta_{i,l,k}}{\bm \psi}_l+{\bf Z}_{B_i},
\end{eqnarray}
where $e^{j\theta_{i,l,k}}$ is the effectively shifted phase from the $k$th user in the $l$th cell to BS $i$. It can be easily to see that such phase offset can be absorbed in the instantaneous fading coefficient and hence does not affect the system performance.

Next, we analyze the case that the time difference between the pilot arrivals at the BS is considerable
compared with one symbol duration. We define the pilot sequence stream function as below
\begin{eqnarray}
\psi_i(m)=\psi_{im},~m=1,...,\omega
\end{eqnarray}
where $\psi_i(m)$, the $m$th element of the pilot sequence ${\bm \psi}_i$, denotes the pilot symbol transmitted by the users of cell $i$ at time $m$. Let $\tau_{i,l,k}$ denote the propagation delay from BS $i$ to the $k$th user in the $l$th cell. Further let $A(t)$ denote a unit-energy baseband rectangular pulse of duration $T_p$ used by the users. Based on the signal model in \cite{Zhanghongyuan}, we can rewrite the equivalent received signal of BS $i$ at time $t$ during the pilot transmission stage as follows
\begin{eqnarray}\nonumber
{\bf r}_{B_i}(t)~~~~~~~~~~~~~~~~~~~~~~~~~~~~~~~~~~~~~~~~~~~~~~~~~~~~~~~~~~~~\\\nonumber=\sum\limits_{m=1}^\omega\sum\limits_{l=1}^N
\sum\limits_{k=1}^K\sqrt{p_{l,k}\omega}A\left[t-(m-1)T_p-\tau_{i,l,k}
\right]{\bf g}_{i,l,k}\psi_l(m)\\+{\bf z}_{B_i}(t)~~
\end{eqnarray}
where ${\bf z}_{B_i}(t)$ is the additive white Gaussian noise vector.

At BS $i$, the received signal at time $t$, ${\bf r}_{B_i}(t)$, is passed through a filter matched to $A\left[t-(m-1)T_p-\bar{\tau}_{B_i}\right]$, where $\bar{\tau}_{B_i}$ can be chosen as the average delay from the users to BS $i$ without loss of generality. Then, we obtain that
\begin{eqnarray}
{\bf y}_{B_i}(m)=\sum\limits_{l=1}^N\sum\limits_{k=1}^K\sqrt{p_{l,k}\omega}{\bf g}_{i,l,k}\tilde{\psi}_{i,l,k}(m)+{\bf z}_{B_i}(m)
\end{eqnarray}
where $\tilde{\psi}_{i,l,k}(m)$ is the received pilot symbol from the $k$th user in the $l$th cell at time $m$, which is polluted by asynchrony. It can be seen that $\tilde{\psi}_{i,l,k}(m)$ is determined by the time difference $\tau_{i,l,k}-\bar{\tau}_{B_i}\neq 0$. Let $\delta_{i,l,k}\triangleq (\tau_{i,l,k}-\bar{\tau}_{B_i}) \mod T_p$ and the resulting quotient be $m_{i,l,k}$, we have
\begin{eqnarray}\nonumber
\tilde{\psi}_{i,l,k}(m)=\rho(\delta_{i,l,k})\psi_l(m+m_{i,l,k})~~~~~~~~~~~~~~~\\+\rho(T_p-\delta_{i,l,k})\psi_l(m+m_{i,l,k}-1)
\end{eqnarray}
where $\rho(\tau)=\int_0^{T_p}A(t)A(t-\tau)dt$. We can see that the polluted pilot symbol $\tilde{\psi}_{i,l,k}(m)$ is the linear combination of two consecutive symbols $\psi_l(m+m_{i,l,k})$ and $\psi_l(m+m_{i,l,k}-1)$. Collecting the received symbols for all the $\omega$ transmission intervals at BS $i$, we have
\begin{eqnarray}
{\bf Y}_{B_i}=\sum\limits_{l=1}^N\sum\limits_{k=1}^K\sqrt{p_{l,k}\omega}{\bf g}_{i,l,k}\tilde{{\bm \psi}}_{i,l,k}+{\bf Z}_{B_i}
\end{eqnarray}
where $\tilde{{\bm \psi}}_{i,l,k}=\left[\tilde{\psi}_{i,l,k}(1),...,\tilde{\psi}_{i,l,k}(\omega)\right]$ is the received polluted pilot sequence.

As in the perfect synchronization case, BS $i$ tries to estimate the composite channel for the users in cell $i$ as
\begin{eqnarray}\nonumber
{\bf \hat{g}}_{i,c}={\bf Y}_{B_i}{\bm \psi}_i^H~~~~~~~~~~~~~~~~~~~~~~~~~~~~~~~~~~~~~~~~~~~~~~~~~~\\\nonumber=\sum\limits_{k=1}^K\sqrt{p_{i,k}\omega}{\bf g}_{i,i,k}\tilde{{\bm \psi}}_{i,i,k}{\bm \psi}_i^H~~~~~~~~~~~~~~~~~~~~~~~~~~~~~~~~~~\\\nonumber+\sum\limits_{l\neq i}^N\sum\limits_{k=1}^K\sqrt{p_{l,k}\omega}{\bf g}_{i,l,k}\tilde{{\bm \psi}}_{i,l,k}{\bm \psi}_i^H+{\bf z}_{B_i}\\=\underbrace{\sum\limits_{k=1}^K\sqrt{p_{i,k}\omega}\kappa_{i,i,k}{\bf g}_{i,i,k}}_{\mbox{desired}~ \mbox{CSI}}+\underbrace{\sum\limits_{l\neq i}^N\sum\limits_{k=1}^K\sqrt{p_{l,k}\omega}\kappa_{i,l,k}{\bf g}_{i,l,k}}_{\mbox{undesired}~ \mbox{CSI}}+{\bf z}_{B_i}
\end{eqnarray}
where $\kappa_{i,i,k}=\tilde{{\bm \psi}}_{i,i,k}{\bm \psi}_i^H$ and $\kappa_{i,l,k}=\tilde{{\bm \psi}}_{i,l,k}{\bm \psi}_i^H$. Since $\tilde{{\bm \psi}}_{i,l,k}\neq {\bm \psi}_l$, the orthogonality of the pilot sequences is destroyed, which implies that $\kappa_{i,i,k}\neq 1$ and $\kappa_{i,l,k}\neq 0,~\forall l\neq i$. Thus, compared with the perfect synchronous case, the estimated composite channel suffers a \emph{scaling loss} as well as \emph{pilot contamination}.

Finally, we demonstrate the asymptotical SINR of each user when the above mismatched channel is used in the beamformer design. Specifically, BS $i$ applies the following normalized beamformer
\begin{eqnarray}\nonumber
{\bf w}_i=\frac{{\bf \hat{g}}_{i,c}}{\|{\bf \hat{g}}_{i,c}\|}~~~~~~~~~~~~~~~~~~~~~~~~~~~~~~~~~\\\nonumber=\frac{\sum\limits_{l=1}^N\sum\limits_{k=1}^K\sqrt{\omega\beta_{i,l,k}p_{l,k}}\kappa_{i,l,k}{\bf h}_{i,l,k}+{\bf z}_{B_i}}{\mu_i\sqrt{M}}
\end{eqnarray}
where $\mu_i=\frac{\|{\bf \hat{g}}_{i,c}\|}{\sqrt{M}}$ is the normalization factor. Similar to the analysis with pilot contamination in Section IV, we can compute the asymptotic SINR of each user for the non-orthogonality scenario as
\begin{eqnarray}\nonumber
\lim\limits_{M\rightarrow \infty}\mbox{SINR}_{i,k}=\frac{\frac{E_i}{\mu_{i,\infty}^2}\beta_{i,i,k}^2\kappa_{i,i,k}^2}{\sum\limits_{j\neq i}^N\frac{E_j}{\mu_{j,\infty}^2}\beta_{j,i,k}^2\kappa_{j,i,k}^2+\frac{\sigma^2}{\omega p_{i,k}}}\triangleq \mathcal{B}
\end{eqnarray}
where $\mu_{i,\infty}^2\triangleq\lim\limits_{M\rightarrow \infty}\mu_i^2$. When $E_i=E\rightarrow \infty,\forall i$, we have
\begin{eqnarray}\nonumber
\lim\limits_{E\rightarrow \infty}\mathcal{B}=\frac{\beta_{i,i,k}^2\kappa_{i,i,k}^2}{\sum\limits_{j\neq i}^N\frac{\mu_{i,\infty}^2}{\mu_{j,\infty}^2}\beta_{j,i,k}^2\kappa_{j,i,k}^2},
\end{eqnarray}
which reaches a ceiling as in the pilot contamination case.

From the above analysis, we conclude that if the time difference between the pilot arrivals at the BS is
not negligible compared with one symbol duration, the composite channel estimation suffers both scaling loss and pilot contamination and, therefore, the performance of the proposed pilot scheme will be degraded. How to overcome the effect of the non-orthogonality by asynchrony will be left for the future work.

\section{Numerical and Simulation Results}

In this section, we provide numerical and simulation results to illustrate the asymptotic as well as finite-size performance of the considered multicell multicast network with massive MIMO. The cells of the network are hexagonal with a radius (from center to vertex) of $r=1000$ meters. The mobile users are randomly distributed in each cell with uniform distribution, excluding an inner circle of $r_0=100$ meters. The channel bandwidth is $20M$Hz. For the large-scale fading, the distance-dependent path loss in dB is modeled as $PL_{NLOS}=128.1+37.6\log_{10}(d)$, where $d$ is the distance from the user to the BS in kilometers. The log-normal shadowing is considered with $\sigma_{shadow}=8\mbox{dB}$ and the penetration loss is assumed to be $20\mbox{dB}$. The small-scale fading is assumed as the normalized Rayleigh fading. The noise power spectral density is set to be $-174\mbox{dBm}/\mbox{Hz}$, and we set $\sigma_p^2=0.1\sigma^2$. Unless otherwise specified, throughout this section we consider the network containing $N=7$ cells and choose the cell in the center for performance evaluation, which also represents the worst scenario. For each analytical result, $1000$ independent large-scale channel realizations are generated. For simulations, the results for our proposed algorithm are obtained by averaging over $1000$ independent small-scale fading parameters for each large-scale channel realization.

\subsection{Asymptotic SINR Performance with Perfect CSI}

\begin{figure}[tb]
\begin{centering}
\includegraphics[scale=0.45]{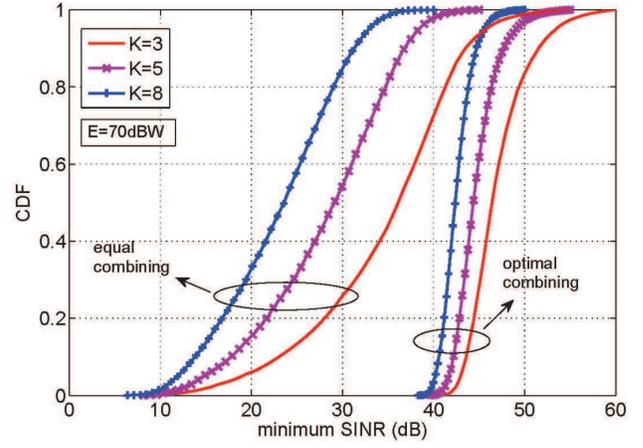}
\vspace{-0.1cm} \caption{The CDF of the minimum  asymptotic SINR of the proposed beamformer under perfect CSI.}
\label{fig:perfect_CSI}
\end{centering}
\vspace{-0.3cm}
\end{figure}

In this subsection, we demonstrate the effectiveness of the asymptotically optimal beamformer when the perfect CSI is available at each BS.

Fig. $\ref{fig:perfect_CSI}$ plots the CDF (Cumulative Distribution Function) curve of the minimum asymptotic SINR among all $K$ users in the central cell. The SINR is computed according to \eqref{asym_SINR1}. For the ``optimal combining" case, the parameters $\{\lambda_{i,k}\}$ are set as in \eqref{th3_2}; For the ``equal combining" case, the parameters $\{\lambda_{i,k}\}$ are set as in \eqref{th2_1} with $\{\xi_{i,k}\}$ being $1$. We can see that by using the beamformer with the optimal combination coefficients, the $50$-th percentile of SINR of the system can be improved by $10$ dB for $K=3$  over that with equal combination coefficients, and the gain increases when $K$ increases. This clearly shows the significance of the linear combination coefficients in the proposed asymptotically optimal beamformer structure. From Fig. $\ref{fig:perfect_CSI}$ we also see that the minimum asymptotic SINR decreases when the user number increases.

\subsection{Asymptotic SINR Performance with Imperfect CSI}

\begin{figure}[tb]
\begin{centering}
\includegraphics[scale=0.45]{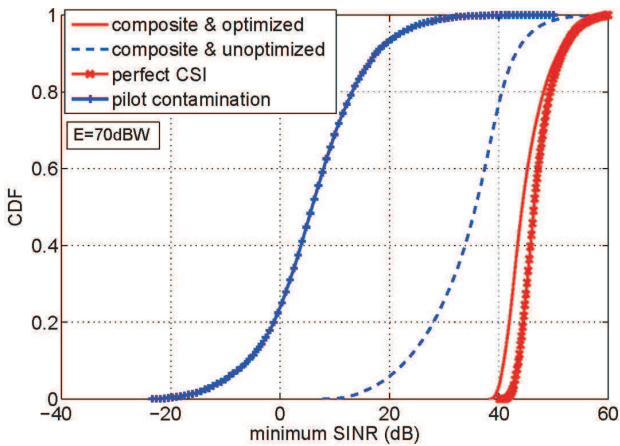}
\vspace{-0.1cm} \caption{The CDF of the minimum asymptotic SINR of the proposed beamformer with different CSI knowledge at $K=3$ users per cell.}
\label{fig:composite_N7k3_CDF}
\end{centering}
\vspace{-0.3cm}
\end{figure}

\begin{figure}
\begin{centering}
\includegraphics[scale=0.45]{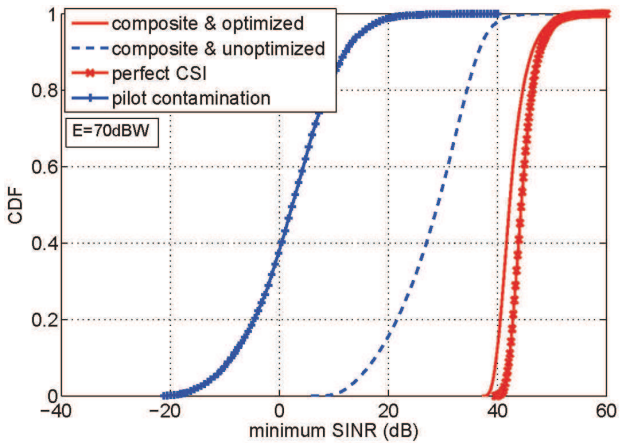}
\vspace{-0.1cm} \caption{The CDF of the minimum asymptotic SINR of the proposed beamformer with different CSI knowledge at $K=5$ users per cell.}
\label{fig:composite_N7k5_CDF}
\end{centering}
\vspace{-0.3cm}
\end{figure}
For the imperfect CSI scenario, each BS obtains the CSI through channel estimation. The length of the pilot sequence is set to be $8$. The pilot peak transmit power is assumed as $p_u=2\mbox{dB}W$ unless otherwise stated. We first investigate in Fig. $\ref{fig:composite_N7k3_CDF}$ and Fig. $\ref{fig:composite_N7k5_CDF}$ the asymptotic performance of the proposed beamformer with different CSI knowledge: the ideal case with perfect CSI, the conventional individual channel estimation with pilot contamination, and the proposed composite channel estimation using contamination-free pilot. The asymptotic SINR of the ``pilot contamination" case is computed according to \eqref{SINR_contamination} and we set $\xi_{i,k}=1, \forall i,\forall k$ since we do not consider parameter optimization. The asymptotic SINR of the composite channel estimation is computed according to (36). For the ``composite $\&$ optimized" case, the pilot power is set according to \eqref{simplified}; for the ``composite $\&$ unoptimized" case, the pilot power of each user is set to be $p_u$. It is seen that the proposed composite channel estimation scheme can benefit significantly from the optimal pilot power control. In particular, the performance with optimal power control is close to the perfect CSI case (with optimal combining coefficients).
It is also seen that the performance of the proposed pilot scheme for composite channel estimation without pilot power control is still much better than that of the conventional individual channel estimation scheme.

\begin{figure}
\begin{centering}
\includegraphics[scale=0.45]{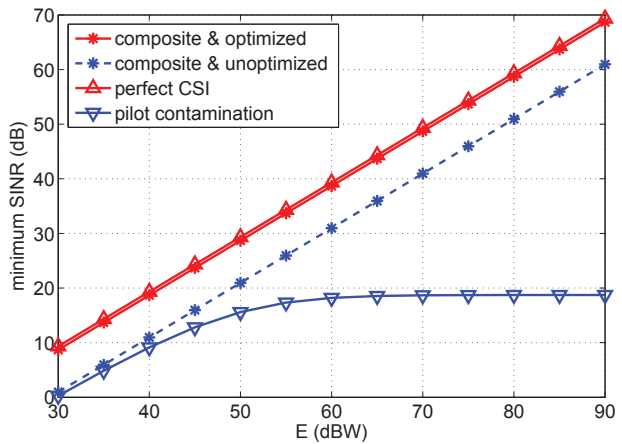}
\vspace{-0.1cm} \caption{The averaged minimum asymptotic SINR of the proposed beamformer with different CSI knowledge at $K=3$ users per cell with varying $E$.}
\label{fig:composite_N7K3}
\end{centering}
\vspace{-0.3cm}
\end{figure}

\begin{figure}
\begin{centering}
\includegraphics[scale=0.45]{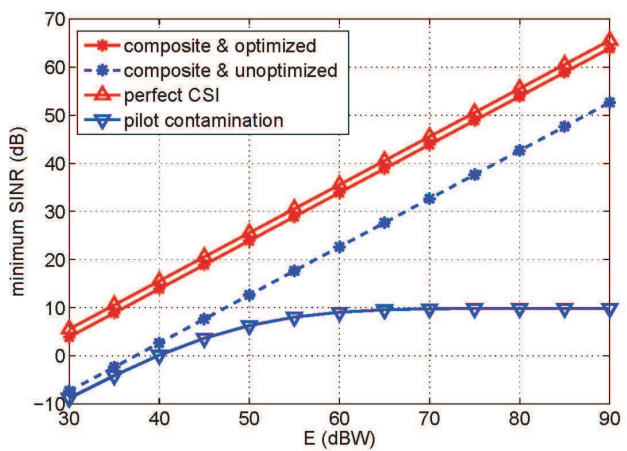}
\vspace{-0.1cm} \caption{The averaged minimum asymptotic SINR of the proposed beamformer with different CSI knowledge at $K=5$ users per cell with varying $E$.}
\label{fig:composite_N7K5}
\end{centering}
\vspace{-0.3cm}
\end{figure}

In Fig. $\ref{fig:composite_N7K3}$ and Fig. $\ref{fig:composite_N7K5}$, we plot the asymptotic SINR curves for different $E$'s and similar phenomenon can be seen. We also observe that when increasing $E$, the average mininum SINR increases almost linearly. However, for the conventional individual channel estimation scheme, the asymptotic SINR reaches a ceiling when $E$ becomes large due to the pilot contamination.

\begin{figure}
\begin{centering}
\includegraphics[scale=0.45]{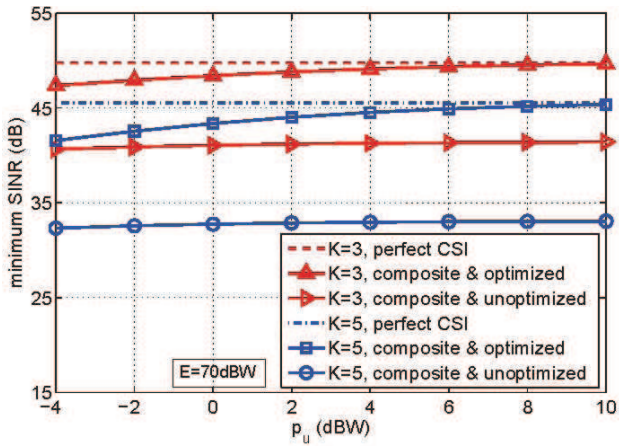}
\vspace{-0.1cm} \caption{The performance the proposed composite channel estimation with different pilot power levels.}
\label{fig:composite_different_pu}
\end{centering}
\vspace{-0.3cm}
\end{figure}

Next, we consider the effects of maximum pilot transmit power on the proposed composite channel scheme. In Fig. $\ref{fig:composite_different_pu}$, we plot the asymptotic SINR of the proposed pilot scheme for different levels of pilot transmitting power. It can be seen that its performance is closer to the perfect CSI case as the pilot transmitting power increases. This is because larger pilot transmitting power leads to smaller effect of the noise in the channel estimation.

\subsection{Actual SINR Performance at Finite Number of Antennas}
In this subsection, we show the actual SINR performance of the proposed scheme by simulation at finite number of antennas. We first compare the performance of the asymptotically optimal multicast beamformer in (\ref{w_i1}) with the traditional SDR-based multicast beamformer \cite{Sidiropoulos,Karipidis,Gaoyan,Xiang,Ozbek}\footnote{For the SDR based algorithm, due to its high complexity, we generate $100$ independent large-scale channel realizations and $100$ independent small-scale fading parameters for each large-scale channel realization to maintain moderate calculation works in the simulation.} when perfect CSI is available. Then, we show the performance when using the new pilot scheme for composite channel estimation.

\begin{figure}
\begin{centering}
\includegraphics[scale=0.45]{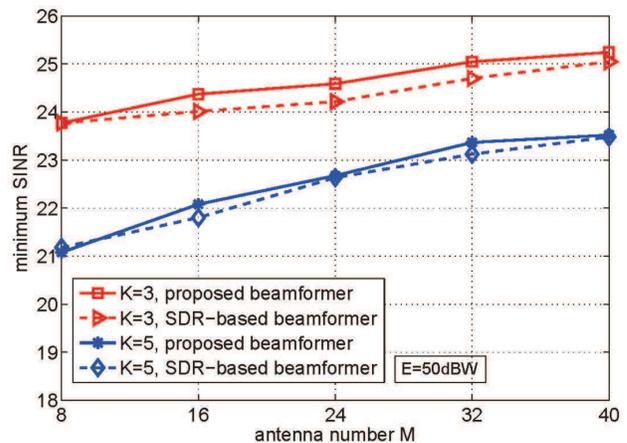}
\vspace{-0.1cm} \caption{Performance comparison between the proposed beamformer and the noncooperative SDR-based beamformer with perfect CSI.}
\label{fig:proposed_vs_SDR}
\end{centering}
\vspace{-0.3cm}
\end{figure}

 In the perfect CSI case, we first compare with the noncooperative SDR-based algorithm in which each BS individually designs their multicast beamformer to serve its users using its local CSI, and treats the intercell interference as noise. Fig.~\ref{fig:proposed_vs_SDR} shows the averaged minimum SINR performance. We can see that our proposed beamformer slightly outperforms the noncooperative SDR beamformer with the considered number of antennas. However, the complexity of the noncooperative SDR algorithm is much higher since it uses interior-point method to iteratively solve the semidefinite programming problem and applies bisection method for the max-min fairness design criterion. Our proposed beamformer, on the other hand, has closed-form solution and no numerical computation is needed.

\begin{figure}
\begin{centering}
\includegraphics[scale=0.45]{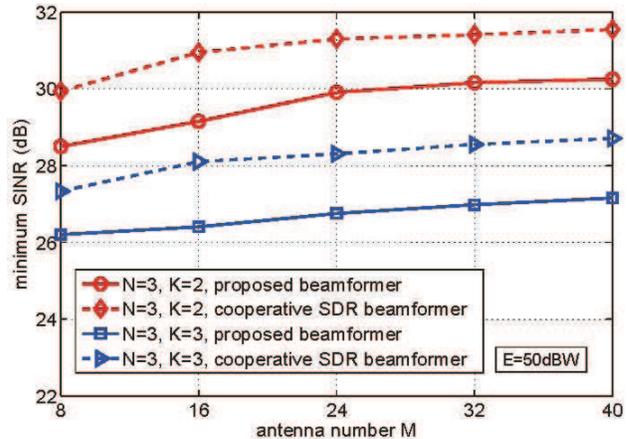}
\vspace{-0.1cm} \caption{Performance comparison between the proposed beamformer and the cooperative SDR-based beamformer with perfect CSI at $N=3$ cells.}
\label{fig:proposed_vs_cooperativeSDR}
\end{centering}
\vspace{-0.3cm}
\end{figure}

 We then compare with the cooperative SDR-based algorithm in which all the BSs coordinate with each other to design their multicast beamformers and the global CSI is needed for each BS. Note that the cooperative SDR beamformer has both much higher computational complexity and much higher signaling overhead. To lighten the computation burden of simulation, we consider the $N=3$ case instead of $N=7$ case in this comparison.  The results are shown in Fig. \ref{fig:proposed_vs_cooperativeSDR}. It can be seen that the cooperative SDR beamformer outperforms our proposed beamformer with the considered number of antennas. This is reasonable since the cooperative SDR beamformer can be near-optimal for small number of users \cite{Xiang}. However, the cooperative SDR beamformer can hardly be used in the cellular network with massive MIMO since the computation complexity is too high and the global CSI is also very hard to be obtained at the BS end. Our proposed multicast beamformer is asymptotically optimal and only needs local CSI.

\begin{figure}
\begin{centering}
\includegraphics[scale=0.45]{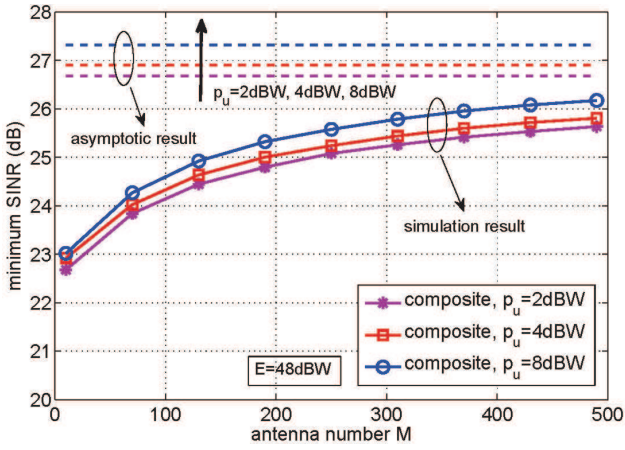}
\vspace{-0.1cm} \caption{The averaged minimum SINR of the proposed pilot scheme for composite channel estimation with finite $M$ and at $K=3$ users per cell.}
\label{fig:simulated_pu248}
\end{centering}
\vspace{-0.3cm}
\end{figure}

When the proposed pilot scheme for composite channel estimation is applied, the beamformer is obtained according to (\ref{composite_w}), where the pilot power is optimized according to (\ref{simplified}).
 Fig. $\ref{fig:simulated_pu248}$ shows the averaged minimum SINR of our proposed scheme at finite antenna number $M$ and with different peak pilot power $p_u$. For comparison the asymptotic SINR obtained using \eqref{asym_SINR2} at infinite $M$ is also plotted. It is seen that when $M=300$, the actual achieved minimum SINR is only $1.5$dB worse that the asymptotic result. The performance gap reduces to $1$dB when $M=500$.

\section{Conclusions}
This paper considered the noncooperative multicell multicast network with massive MIMO. For the perfect CSI case, we obtained the asymptotically optimal beamformer for each BS in closed-form that is a function of the large-scale channel attenuation $\{\beta_{i,j,k}\}$. For the imperfect CSI, we first considered the conventional individual channel estimation scheme and analyzed the effect of the pilot contamination. Then we proposed a novel pilot scheme for composite channel estimation that completely eliminates pilot contamination. By performing power control among users in the pilot transmission stage, our proposed scheme offers performance that is close to the performance under the perfect CSI scenario. Besides, we analyzed the effect of asynchrony on the performance of our proposed scheme.

\appendices
\section{Proof of Lemma 1}
We first prove \eqref{th3_1} by contradiction. Let $\{\lambda_{i,k}^*\}$ be the optimal solution of problem ${\cal Q}$ but the condition \eqref{th3_1} does not hold. Without loss of generality, we assume $\frac{\lambda_{i,1}^*E_i\beta_{i,i,1}}{\sigma^2}$ is the exclusive maximum one and $\frac{\lambda_{i,K}^*E_i\beta_{i,i,K}}{\sigma^2}$ is the exclusive minimum one\footnote{The extension to the cases that there are multiple equal maximum SINR's and multiple equal minimum SINR's is straightforward and is omitted here.}, and $\frac{\lambda_{i,1}^*E_i\beta_{i,i,1}}{\sigma^2}>\frac{\lambda_{i,K}^*E_i\beta_{i,i,K}}{\sigma^2}$. We can then reset $\lambda_{i,1}^*$ and $\lambda_{i,K}^*$ as follows
\begin{eqnarray}\nonumber
{\tilde{\lambda}_{i,1}^*}&=&\lambda_{i,1}^*-\triangle \lambda,\\
{\tilde{\lambda}_{i,K}^*}&=&\lambda_{i,K}^*+\triangle \lambda, \nonumber
\end{eqnarray}
where
\begin{eqnarray}\nonumber
\triangle\lambda=\frac{\lambda_{i,1}^*\beta_{i,i,1}-\lambda_{i,K}^*\beta_{i,i,K}}{\beta_{i,i,1}+\beta_{i,i,K}}>0.
\end{eqnarray}
With the new parameters, we have that
\begin{eqnarray}\nonumber
\frac{\tilde{\lambda}_{i,1}^*E_i\beta_{i,i,1}}{\sigma^2}=\frac{\tilde{\lambda}_{i,K}^*E_i\beta_{i,i,K}}{\sigma^2}>\frac{\lambda_{i,K}^*E_i\beta_{i,i,K}}{\sigma^2}.
\end{eqnarray}
It means that the objective value of ${\cal Q}$ has been increased, which contradicts to the assumption that $\{\lambda_{i,k}^*\}$ is the optimal solution of problem ${\cal Q}$. Thus, we must have condition \eqref{th3_1}.

Combining \eqref{ob_lamda} and \eqref{th3_1}, we can easily obtain \eqref{th3_2} and thus the lemma is proven.

\section{Proof of Lemma 2}
We prove \eqref{p_star} by contradiction. If $p_{i,k^\star}<p_u$, we will have that
\begin{eqnarray}\label{p_lemma2}
p_{i,k}<p_u, ~\forall k.
\end{eqnarray}
Supposing \eqref{p_lemma2} does not hold, we can assume $p_{i,1}=p_u$ without loss of generality. Since $\beta_{i,i,1}^2p_{i,1}>\beta_{i,i,k^\star}^2p_{i,k^\star}$ and the expression of each user's SINR has a common denominator $\sum\limits_{k'=1}^K\beta_{i,i,k'}p_{i,k'}+\frac{\sigma_p^2}{\omega}$, user $1$'s SINR is larger than that of user $k^\star$. Then, we can set user $1$'s pilot power as $p'_{i,1}<p_u$ to make that $\beta_{i,i,1}^2p'_{i,1}=\beta_{i,i,k^\star}^2p_{i,k^\star}$. Thus, the objective value of problem \eqref{Ti_ob1} will increase since the denominator $\sum\limits_{k'=1}^K\beta_{i,i,k'}p_{i,k'}+\frac{\sigma_p^2}{\omega}$ decreases, which contradicts with the optimality.

Based on \eqref{p_lemma2}, we can always increase each user's pilot power as $p'_{i,k}=\eta p_{i,k},\forall k$ where $\eta>1$ without violating the power constraints. Then each item in the objective function of problem \eqref{Ti_ob1} will be
\begin{eqnarray}
\frac{\beta_{i,i,k}^2\eta p_{i,k}}{\sum\limits_{k'=1}^K\beta_{i,i,k'}\eta p_{i,k'}+\frac{\sigma_p^2}{\omega}}=\frac{\beta_{i,i,k}^2p_{i,k}}{\sum\limits_{k'=1}^K\beta_{i,i,k'}p_{i,k'}+\frac{\sigma_p^2}{\eta\omega}}
\end{eqnarray}
which increases with $\eta>1$. Then the objective value of problem \eqref{Ti_ob1} will increase which contradicts with the optimality. Thus, we must have \eqref{p_star}.

\bibliographystyle{IEEEtran}
\bibliography{IEEEabrv,multicast_massive_MIMO}

\end{document}